\documentclass[IEEEjournal]{IEEEtran}
\usepackage{graphicx}
\usepackage{subfigure}
\usepackage{algorithm}
\usepackage{algorithmic}
\usepackage{cite}
\usepackage{amsfonts}
\usepackage{amssymb}
\usepackage{psfrag}
\usepackage{footnote}
\usepackage{amsmath}
\usepackage[switch,pagewise]{lineno}
\usepackage{stfloats}
\usepackage{rotating}

\newtheorem{lemma}{\textbf{Lemma}}
\newtheorem{theorem}{\textbf{Theorem}}

\begin{document}
%
%
\title{On Pollaczek-Khinchine Formula for Peer-to-Peer Networks}

\author{Jian Zhang,
		Tony T. Lee,~\IEEEmembership{Fellow,~IEEE,}
		Tong Ye,~\IEEEmembership{Member,~IEEE,}
        and Weisheng Hu,~\IEEEmembership{Member,~IEEE}
\thanks{This work was supported by the National Science Foundation of China (61271215 and 61571288)}
\thanks{The authors are with the State Key Laboratory of Advanced Optical Communication
Systems and Networks, Shanghai Jiao Tong University, Shanghai
200030, China. (e-mail: \{2011aad, ttlee, yetong, wshu\}@sjtu.edu.cn)}
}

\maketitle

\begin{abstract}
The performance analysis of peer-to-peer (P2P) networks calls for a new kind of queueing model, in which jobs and service stations arrive randomly. Except in some simple special cases, in general, the queueing model with varying service rate is mathematically intractable. Motivated by the P-K formula for M/G/1 queue, we developed a limiting analysis approach based on the connection between the fluctuation of service rate and the mean queue length. Considering the two extreme service rates, we proved the conjecture on the lower bound and upper bound of mean queue length previously postulated. Furthermore, an approximate P-K formula to estimate the mean queue length is derived from the convex combination of these two bounds and the conditional mean queue length under the overload condition. We confirmed the accuracy of our approximation by extensive simulation studies with different system parameters. We also verified that all limiting cases of the system behavior are consistent with the predictions of our formula. 
\end{abstract}


\IEEEpeerreviewmaketitle

\section{Introduction}\label{intro}
\IEEEPARstart Recently, applications of Peer-to-Peer (P2P) systems have been widely deployed in the internet. The studies reported in \cite{wehrlepeer} reveal that more than 50\% of internet traffic, sometimes even more than 75\%, is attributed to P2P applications. The P2P systems provide scalable and reliable high throughput service because each client is also a server. That is, when a client receives service, it also provides service to other clients. The two major P2P applications are video streaming and online file sharing \cite{wehrlepeer}. The video streaming system uses P2P to share bandwidth and achieve load-balancing among clients. It aims at bandwidth allocation among peers to guarantee that every client can receive service in time with high probability \cite{Wu2009}\cite{KumarP2Pstreaming}. On the other hand, the concern of online file sharing systems is about file transfer delay within the network \cite{qiu2004modeling}\cite{FanBitTorrent}\cite{de2003fairness}. In this paper, we focus on the delay analysis of P2P systems.

The performance analysis of P2P systems spurs many to embark on a new kind of queueing modeling \cite{li2009queuing}\cite{Taoyujournal}, in which jobs and servers arrive randomly. In queueing theory \cite{kleinrock1975theory}, the system with a dynamic number of servers can be regarded as a single server queue with a variable service rate, which is a hot topic arising in several diversified research fields, such as multi-rate wireless channels \cite{HuangLiangGPK}\cite{Twostate2010} and mobile cloud computing \cite{Mobile2014}. In many published works, the server with several service rates is described as a continuous-time Markov chain with a finite number of states \cite{HuangLiangGPK}\cite{Twostate2010}\cite{mahabhashyam2005queues}\cite{eisen1963stochastic}. 

The major obstacle to analyzing the queueing model with varying service rate is the dependency on service times among different customers. To cope with this issue, Mahabhashyam and Gautam introduced the concept of start service probability in \cite{mahabhashyam2005queues}. Huang and Lee adopted this concept in \cite{HuangLiangGPK} and derived the generalized P-K formula for two-state Markov channels. Despite that the closed-form mean queue length formula can be obtained for some special cases, the P2P queueing model is mathematically intractable when using this kind of method because the server of a P2P system may have an infinite number of service rates.

\subsection{Previous work}\label{pre-work}

The first model of P2P systems was proposed by Qiu and Srikant in \cite{qiu2004modeling} to study BitTorrent, a P2P file sharing system. With the assumption of random arrival and departure of service stations, they analyzed the queueing delay and service time of each request. Although the analysis is simple, their model attracts much attention. For example, the Qiu-Srikant model was later generalized by Clevenot, Nain and Ross to more realistic P2P systems in \cite{clevenot2005multiclass}, where peers can be non-homogeneous. The main concern of video streaming is the probability of receiving service in time. A stochastic fluid model was adopted by the authors of \cite{Wu2009} and \cite{KumarP2Pstreaming} to analyze this aspect of P2P streaming systems.

The queuing model with variable service rate is also an important analytical tool in studying multi-rate wireless channels. Eisen and Tainiter first introduced the two-state queueing model in their 1963 paper \cite{eisen1963stochastic}. Gunaseelan, Liu and Chamberland \cite{Twostate2010} derived the mean delay for the system with a two-state server, in which one of the service rates is 0. To cope with the dependencies among service times, a novel approach based on conditional moments of service time is proposed by Mahabhashyam and Gautam in \cite{mahabhashyam2005queues}. However, their analysis is incomplete because the required start-service probabilities are only available for some extreme cases. The complete start service probabilities and the closed-form mean delay formula for general two-state queueing model were derived by Huang and Lee in \cite{HuangLiangGPK}. 

The P2P queueing model proposed in \cite{li2009queuing} and \cite{Taoyujournal} is similar to that of the two-state model of wireless channels. The only difference is that the Markov chain of the P2P model has an infinite number of states, and the server process can be viewed as an $M/M/\infty$ queue. In \cite{Taoyujournal}, Taoyu Li et al. proposed two bounds of mean queueing delay, but their proof is based on a conjecture that cannot be verified.

\subsection{Our approach and contribution}\label{OurApproach}

In this paper, based on the connection between the fluctuation of service rate and the mean queue length, we provide a rigorous proof for the two bounds conjectured in \cite{Taoyujournal}, and derive an approximate Pollaczek-Khinchine formula of mean queue length for P2P networks. Our methodology and results are summarized as follows:

\begin{enumerate}
  \item We prove that the mean and variance of a service time increase with the variance of the service rate if we keep the mean service rate constant. As the P-K formula of M/G/1 queue demonstrates, the mean queue length is strongly influenced by the first two moments of service time. Thus, we establish a relationship between the service rate fluctuation and the mean queue length. 

  \item Based on the relationship between service rate fluctuation and mean queue length, we prove the two bounds postulated in \cite{Taoyujournal} by considering the two limiting cases of service rate. When the variance of service rate approaches 0, the mean queue length reaches the lower bound. When the variance approaches infinite, the mean queue length reaches the upper bound. This result also confirms the conjecture that the number of servers and the number of jobs in a P2P system are negatively correlated.

  \item From the convex combination of the lower bound and upper bound, we derive an approximate P-K formula of the mean queue length for P2P networks. The accuracy of our approximation has been verified by extensive simulations, and all limiting cases of the P2P system behavior agree with the predictions made by the formula.

\end{enumerate}

The limiting analysis and approximate estimation of mean queue length developed for the P2P networks can be extended and applied to other queueing systems with varying service rates. The innovative approaches proposed in this paper that are of particular interest include the extreme analysis of the service rate fluctuations, and the technique of approximate estimation based on the conditional mean queue length and convex combination of extreme bounds.

The rest of the paper is organized as follows. In Section \ref{queueing_model}, we describe the P2P queueing model and derive the differential equation associated with the number of jobs and number of servers in the system. A simple relationship between these two numbers is obtained from the differential equation. In Section \ref{Bounds}, we demonstrate the connection between the service rate fluctuation and the mean queue length, and confirm the conjecture on the two bounds postulated in \cite{Taoyujournal}. In Section \ref{approximation}, based on the conditional mean queue length and the convex combination of the two bounds, we derive an approximate mean queue length formula, and verify the accuracy of our approximation by using extensive simulations and limiting analysis. Section \ref{conclusion} draws a conclusion.

\section{Markov Chain of P2P Networks}\label{queueing_model}

\subsection{System description}\label{system_description}
\begin{figure}[H]
\centering
\includegraphics[width=90mm]{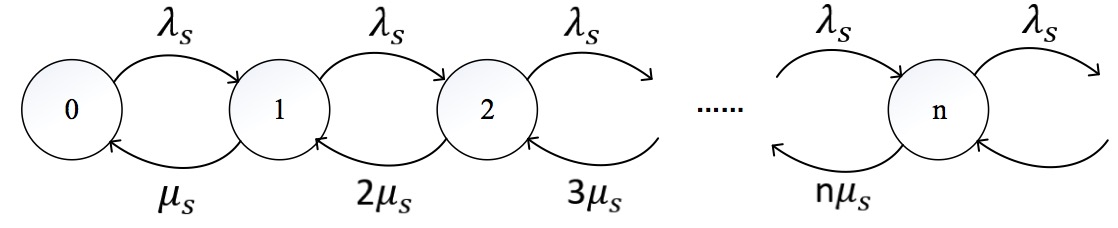}
\caption{The continuous time Markov chain of the server process.}\label{figure1}
\end{figure}

In a P2P service system, jobs and servers may randomly arrive and depart. In the queuing model under consideration, we assume that the server arrival process $\{n_s(t), t \geq 0 \}$ is a Poisson process with rate $\lambda_s$, and servers are homogeneous, meaning independent and identical. The lifetime of a server is exponentially distributed with mean $\frac{1}{\mu_s}$, and the service rate of each server is $\mu_c$  jobs per unit of time. As Fig. \ref{figure1} shows, the continuous time Markov chain of the server process can be viewed as an $M/M/\infty$  queue. Therefore, in steady state, the number of servers in the system, defined as $n_s=\lim\limits_{t\rightarrow{\infty}} n_s(t)$, is a Poisson random variable with parameter $\rho_s = \frac{\lambda_s}{\mu_s}$.

Furthermore, we assume that the job arrival process $\{n_c(t), t \geq 0 \}$ is also a Poisson process with rate $\lambda_c$, which is independent of the server process. The service of jobs waiting in line follows first-come, first-served (FCFS) policy. At any instant of time, only one job will be simultaneously served by all servers. Therefore, the job service time depends on the number of servers in the system. If the number of servers is a constant within the service time of a job, then the job service time is exponentially distributed. However, as the number of servers may change within the service time of a job, the exact job service time distribution is unknown. Since the service times of jobs are dependent, the Kendall's notation of queueing systems cannot be extended to the P2P queueing model, which is characterized by the following set of parameters: 
\begin{itemize}
 \item Job arrival rate: $\lambda_c$, \\
 	   Service rate of one server: $\mu_c$, \\
 	   $\rho_c = \frac{\lambda_c}{\mu_c}$.
 \item Server arrival rate: $\lambda_s$, \\
 	   Mean server lifetime: $\frac{1}{\mu_s}$, \\
 	   $\rho_s = \frac{\lambda_s}{\mu_s}$.
\end{itemize}

Furthermore, we define the following notations used throughout this paper.
\begin{itemize}
 \item $n_s(t)$: number of servers at time $t$, \\
 	   $n_s = \lim\limits_{t\rightarrow{\infty}} n_s(t)$: Poisson random variable of server number with parameter $\rho_s$ when the system is in steady state.

 \item $n_c(t)$: number of jobs at time $t$, \\
 	   $n_c = \lim\limits_{t\rightarrow{\infty}} n_c(t)$: random variable of job number when the system is in steady state.	

 \item $\mu(t)=n_s(t)\mu_c$: instantaneous service rate at time $t$, \\
       $\mu = \lim\limits_{t\rightarrow{\infty}} \mu(t)$: random variable of service rate when system is in steady state. \\
       $\bar{\mu} = E[\mu]$: mean service rate.
\end{itemize}

\subsection{Continuous-time Markov chain}\label{CTMC}

\begin{figure}[H]
\centering
\includegraphics[width=90mm]{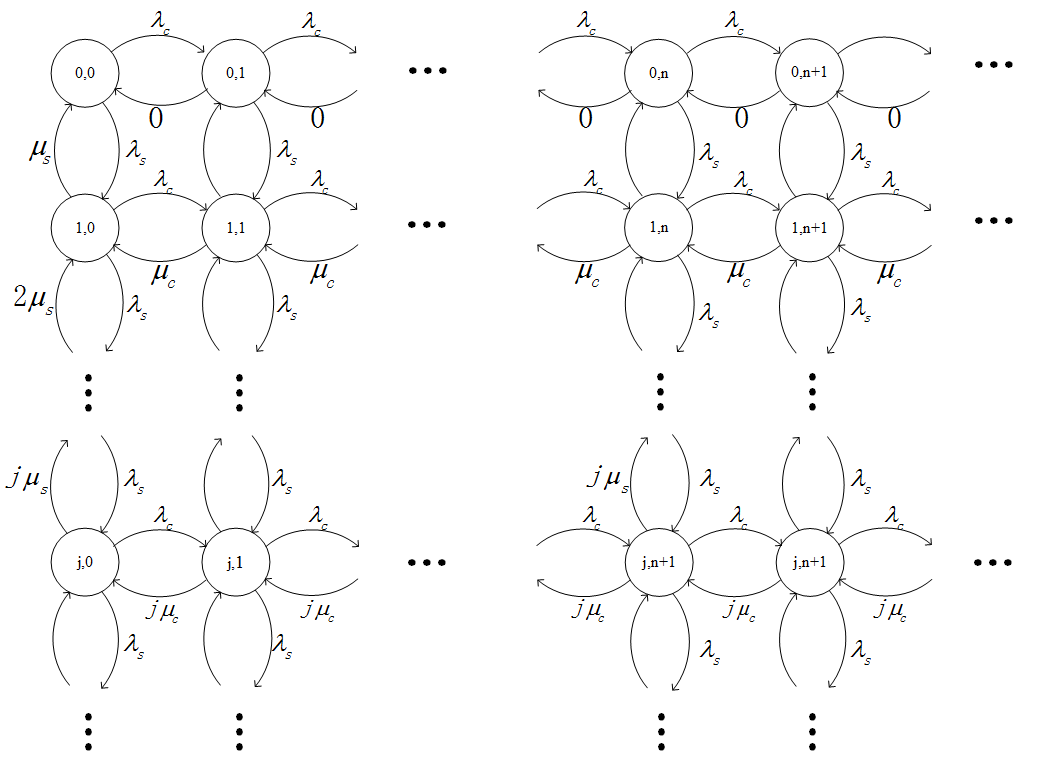}
\caption{The continuous-time Markov chain of the P2P queueing model.}\label{figure2}
\end{figure}

According to the previous P2P system description, Fig. \ref{figure2} shows the continuous-time Markov chain of the P2P queueing model with state space $\{(i,j), i=0,1,2,\dots,j=0,1,2,\dots\}$, where $i$ is the number of jobs and $j$ is the number of servers in the system in steady state.

Let $p_{i,j}$ denote the steady state probability that the system is in state $(i,j)$, that is 
\begin{equation}
 p_{i,j}=Pr\{n_c=i, n_s=j\}, 
\end{equation}
where $i = 0,1,\dots$ is the number of jobs and $j=0,1,\dots$ is the number of servers. From the state transition diagram shown in Fig. \ref{figure2}, we directly obtain the following set of balance equations:
\begin{subequations}\label{balance_equations}
\begin{equation}
 (\lambda_s+\lambda_c)p_{0,0} = \mu_s p_{0,1}     \quad  i=0,j=0,
\end{equation}
\begin{equation}
 (\lambda_s+\lambda_c)p_{i,0} = \lambda_c p_{i-1,0} + \mu_s p_{i,1}   \quad    i\geq1,j=0,
\end{equation}
\begin{eqnarray}
 &&(\lambda_s+\lambda_c+j\mu_s)p_{0,j} = \lambda_s p_{0,j-1} + j\mu_c p_{1,j} \nonumber \\
 && \qquad  + (j+1)\mu_s p_{0,j+1}  \quad i=0,j\geq1,
\end{eqnarray}
\begin{eqnarray}
 (\lambda_s+\lambda_c+j\mu_s+j\mu_c)p_{i,j} = \lambda_s p_{i,j-1} + \lambda_c p_{i-1,j}\nonumber \\
  \quad + j\mu_c p_{i+1,j} + (j+1)\mu_s p_{i,j+1}  \quad  i\geq1,j\geq1.
\end{eqnarray}
\end{subequations}

Define the following generating function of $p_{i,j}$:
\begin{equation}
 F(z_1,z_2) = \sum_{i=0}^{\infty}\sum_{j=0}^{\infty} p_{i,j} z_1^i z_2^j \quad |z_1| \leq 1,|z_2| \leq 1.
\end{equation}

From the set of balance equations (\ref{balance_equations}), we can derive the following differential equation of the generating function $F(z_1,z_2)$:
\begin{eqnarray}\label{eq_of_F}
 && (\lambda_c-z_1\lambda_c+\lambda_s-z_2\lambda_s)F(z_1,z_2) = \nonumber \\
 &&\qquad \quad \mu_s(1-z_2) \frac{\partial F(z_1,z_2)}{\partial z_2} + z_2\mu_c(\frac{1}{z_1}-1)\frac{\partial F(z_1,z_2)}{\partial z_2}\nonumber \\
 &&\qquad \quad +(1-\frac{1}{z_1})z_2\mu_c\frac{\partial F(0,z_2)}{\partial z_2}.
\end{eqnarray}

Unfortunately, to the best of our knowledge, solving this differential equation to obtain a closed-form solution of $F(z_1,z_2)$ is mathematically intractable. That is, we cannot directly obtain the mean queue length of the system from (\ref{eq_of_F}). Nevertheless, this differential equation still provides us with some useful information regarding the performance of P2P system.

First, applying $\frac{\partial}{\partial z_1}$ to equation (\ref{eq_of_F}) and then inserting $z_1 = z_2 =1$, we obtain:
\begin{equation}\label{result1}
 E[n_s] = \rho_c+\frac{\partial F(0,z_2)}{\partial z_2}|_{z_2=1}.
\end{equation}

Similarly, applying $\frac{\partial^2}{\partial z_1^2}$ to equation (\ref{eq_of_F}) and then inserting $z_1 = z_2 =1$, we obtain:
\begin{equation}\label{result2}
 \rho_cE[n_c] = E[n_sn_c]-E[n_s] + \frac{\partial F(0,z_2)}{\partial z_2}|_{z_2=1}.
\end{equation}

Combining (\ref{result1}) and (\ref{result2}), we have
\begin{equation}\label{result3}
 \rho_cE[n_c] = E[n_sn_c]-\rho_c.
\end{equation}

Since the number of jobs $n_c$ is correlated with the number of servers $n_s$, the mean queue length of jobs $E[n_c]$ remains unsolvable from (\ref{result3}). In \cite{Taoyujournal}, Taoyu Li et al. postulated that the number of jobs and the number of servers in the system are negatively correlated, and then proposed a conjecture on the upper bound and the lower bound of the mean queue length of jobs. Intuitively, it seems that the postulation is correct, but there is no rigorous proof available to confirm their conjecture.

\section{Service Rate Fluctuation and Delay Bound}\label{Bounds}
For the P2P queuing model, the lack of service time distribution and the dependency among service times of different jobs are major obstacles for deriving the mean queue length of jobs. The only known information related to the service time is the Poisson distribution of the number of servers, which determines the service rate distribution. In this section, we first investigate the influence of service rate fluctuations on mean queue length of jobs, and then provide a proof of the upper bound and the lower bound of mean queue length conjectured by Taoyu Li et al. in \cite{Taoyujournal}.

\subsection{Influence of service rate fluctuation}\label{service_rate_fluctuation}
Intuitively, both the mean and variance of a service time random variable will increase with respect to the fluctuation of service rate. In the following lemma, we show that this property holds when the mean service rate remains a constant.

\begin{lemma}\label{lemma1}
 If the mean service rate is fixed, then the mean and variance of service time increase with the variance of service rate.
\end{lemma}

\begin{IEEEproof}
 Suppose the service rate is a random variable $\mu$ with mean $E[\mu] = \bar{\mu} \neq 0$. A second-order approximation of a given function $g(\mu)$ by using Taylor series expansion about $\bar{\mu}$ is given by:
 \begin{equation}
  g(\mu) = g(\bar{\mu}) + g'(\bar{\mu})(\mu-\bar{\mu})	+ \frac{g''(\bar{\mu})}{2}(\mu-\bar{\mu})^2 + reminder,
 \end{equation}
 we therefore can obtain the approximations of the mean and variance of  expressed as follows:
 \begin{subequations}
 \begin{equation}\label{mean}
  E[g(\mu)] \cong g(\bar{\mu})	+ \frac{g''(\bar{\mu})}{2}Var[\mu],
 \end{equation}
 \begin{equation}\label{variance}
  Var[g(\mu)] \cong g''(\bar{\mu})^2Var[\mu].
 \end{equation}
 \end{subequations}
 Hence, the mean and variance of service time $S=g(\mu) = \frac{1}{\mu}$ can be approximately estimated by:
 \begin{subequations}
 \begin{equation}\label{mean}
  E[S] = E[g(\mu)] \cong \frac{1}{\bar{\mu}} + \frac{1}{\bar{\mu}^3}Var[\mu],
 \end{equation}
 \begin{equation}\label{variance}
  Var[S] = Var[g(\mu)] \cong \frac{1}{\bar{\mu}^4}Var[\mu].
 \end{equation}
 \end{subequations}
 \end{IEEEproof}

 In a queueing system, for a fixed arrival rate $\lambda$, the mean queue length would monotonically increase with both the mean and variance of the service time $S$. This property can be best illustrated by the following well-known P-K formula of M/G/1 queueing model\cite{kleinrock1975theory}:
 \begin{equation}\label{PK}
  L = \rho + \frac{\rho^2+\lambda^2Var[S]}{2(1-\rho)}.
 \end{equation}

 Since the P-K formula of M/G/1 queue is derived under the assumption that service times are i.i.d. random variables, which does not hold in P2P systems. Thus, substituting the mean $E[S]$ and variance $Var[S]$ of service time given in Lemma 1 directly into the above P-K formula will result in a poor approximation of the mean queue length of jobs.

 From Lemma \ref{lemma1}, however, we know that the mean queue length would increase with the variance of service rate if we keep the mean service rate fixed. Based on this property, we can investigate the limiting cases of mean queue length, and thus confirm the upper bound and the lower bound of mean queue length conjectured in \cite{Taoyujournal}.

\subsection{Delay bounds}\label{delay_bounds}
Given that the number of servers $n_s$ is a Poisson random variable with parameter $\rho_s$, we immediately obtain the following parameters:

\begin{itemize}
 \item Average number of servers: $E[n_s]=\rho_s = \frac{\lambda_s}{\mu_s}$,
 \item Standard deviation of server number: $\sigma[n_s] = \sqrt{\rho_s}$,
 \item Mean service rate: $\bar{\mu} = E[\mu] = \rho_s\mu_c$,
 \item Standard deviation of service rate: $\sigma[\mu] = \mu_c\sqrt{\rho_s} = \sqrt{\bar{\mu}\mu_c}$.
\end{itemize}

To investigate the limiting cases of mean queue length with regard to the service rate fluctuation, according to Lemma \ref{lemma1}, the mean service rate should be kept as a constant. If we keep parameter $\frac{\mu_c}{\mu_s}$ and $\lambda_s$ constant and only change the value of $\mu_c$, then the mean service rate $\bar{\mu}$  would be a constant but the standard deviation $\sigma[\mu] = \sqrt{\bar{\mu}\mu_c}$ increases with $\mu_c$.

\begin{figure}[h]
	\center
    \subfigure[$\mu_c=0.1$]{
    \includegraphics[height=50mm]{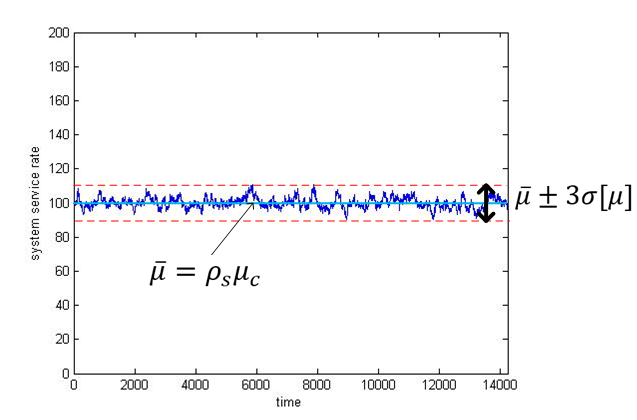}}
    \subfigure[$\mu_c=1$]{
    \includegraphics[height=50mm]{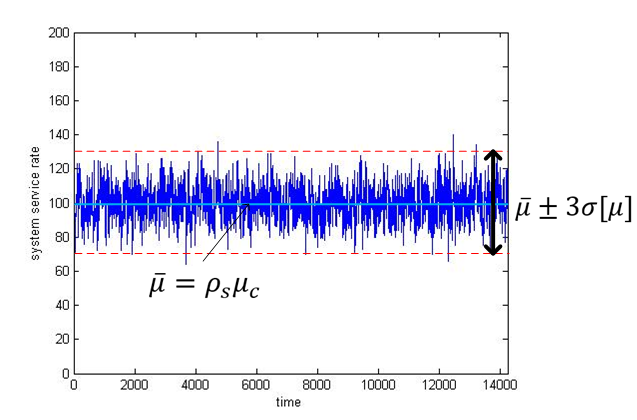}}
    \caption{The fluctuation of service rate $\mu$ over the time with parameter $\frac{\mu_c}{\mu_s} = 10$, $\lambda_s=10$.}\label{figure3}
\end{figure}

For example, if both parameters $\frac{\mu_c}{\mu_s}=10$ and $\lambda_s=10$ are fixed, then the mean service rate  $\bar{\mu}$ equals a constant 100. Fig. \ref{figure3} demonstrates two cases: $\mu_c = 0.1$ in Fig. \ref{figure3}(a) and $\mu_c = 1$ in Fig. \ref{figure3}(b) it is obvious that the service rate fluctuates more in (b) with larger $\mu_c$.  

Intuitively, a larger $\mu_c$ and $\mu_s$ means each server has a relatively large unit service rate and the number of servers is small. In this case, the arrival or departure of a server will lead to a larger change in the service rate than that of a smaller $\mu_c$ and $\mu_s$. When $\mu_c$ and $\mu_s$ are small, each server has a small unit service rate but the number of servers is large. Therefore, an arrival or departure of a single server hardly influences the service rate, which is the reason why the service rate becomes more static.

As the mean queue length increases with the fluctuation of the service rate while keeping both parameters $\frac{\mu_c}{\mu_s}$ and $\lambda_s$ constant, in the following theorem, this property enables us to investigate the two limiting cases of the mean queue length: the lower bound when $\mu_c \rightarrow 0$ and the upper bound when $\mu_c \rightarrow \infty$.

\begin{theorem}\label{theorem1}
 The average queue length $L$ is bounded as follows:
 \begin{subequations}
 \begin{eqnarray}
  &&\text{\textbf{Lower bound: }} L_1 = \frac{\lambda_c}{\bar{\mu}-\lambda_c} \leq L, \\
  &&\text{\textbf{Upper bound: }} L \leq (1+\frac{\mu_c}{\mu_s})\frac{\lambda_c}{\bar{\mu}-\lambda_c} = L_2.
 \end{eqnarray}
 \end{subequations}
\end{theorem}

\begin{IEEEproof}
From Lemma \ref{lemma1}, we know that mean queue length is proportional to the variance of service rate $\mu$, provided that the mean service rate  $\bar{\mu} = \rho_s\mu_c$ is fixed. The standard deviation of service rate is given by $\sigma[\mu] = \sqrt{\bar{\mu}\mu_c}$. For a fixed mean service rate $\bar{\mu}$, we consider two limiting cases, namely $\mu_c \rightarrow 0$ and $\mu_c \rightarrow \infty$.\\

1. Lower bound $L_1 = \lim\limits_{\mu_c,\mu_s\rightarrow{0}}L = \frac{\lambda_c}{\bar{\mu}-\lambda_c}$.\\

For a fixed mean service rate $\bar{\mu} = \rho_s\mu_c = \frac{\lambda_s\mu_c}{\mu_s}$, the limiting case $\mu_c \rightarrow 0$, while keeping both parameters $\frac{\mu_c}{\mu_s}$ and $\lambda_s$ constant, implies $\mu_s \rightarrow 0$ and $\rho_s \rightarrow \infty$. The physical meaning of the P2P system operated under this scenario can be interpreted as follows:

\begin{itemize}
 \item $\rho_s \rightarrow \infty$ implies that the average number of servers is very large,
 \item $\mu_s \rightarrow 0$ implies that the average lifetime of a server $\frac{1}{\mu_s}$ is very long,
 \item $\mu_c \rightarrow 0$ implies that the capacity of each server, in terms of number of jobs served per unit of time, is very small.
\end{itemize}

\begin{figure}[H]
\centering
\includegraphics[height=40mm]{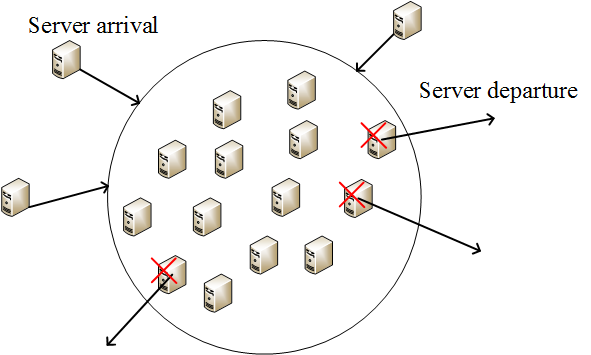}
\caption{Service rate becomes a constant when P2P system reaches equilibrium.}\label{figure4}
\end{figure}

Fig. \ref{figure4} illustrates this scenario. There is a large number of servers, and it seems that these servers always stay in the system because of their extremely long lifetime. However, the service rate (capacity) of each server is very small. According to the law of large numbers, the aggregate service rate $\mu = n_s\mu_c$ should converge to the constant rate $\bar{\mu} = \rho_s\mu_c$. Thus, the system behaves like a single server queue with Poisson arrival rate $\lambda_c$.

Mathematically, when $\mu_c \rightarrow 0$, the standard deviation of service rate $\sigma[\mu] = \sqrt{\bar{\mu}\mu_c}\rightarrow 0$. From Chebyshev's inequality, we have

\begin{equation}
 Pr\{|\mu-\bar{\mu}| \geq \varepsilon\} \leq \frac{E[(\mu-\bar{\mu})^2]}{\varepsilon^2} = \frac{\mu_c \bar{\mu}}{\varepsilon^2} \rightarrow 0,   \quad \text{as } \mu_c \rightarrow 0,
\end{equation}
which means the service rate $\mu = n_s\mu_c$ converges to the mean $\bar{\mu} = \rho_s\mu_c$ with probability 1. It follows that
\begin{equation}
 -\varepsilon E[n_c] \leq E[(\mu-\bar{\mu})n_c] \leq \varepsilon E[n_c], \quad \text{as } \mu_c \rightarrow 0.
\end{equation}
Since the stable condition implies the mean queue length is finite, i.e. $E[n_c]< \infty$ and the parameter $\varepsilon>0$ can be arbitrarily small, therefore
\begin{equation}
 \lim\limits_{\mu_c,\mu_s\rightarrow{0}} E[(\mu-\bar{\mu})n_c] = 0.
\end{equation}
Hence, we have
 \begin{subequations}
 \begin{equation}
  \lim\limits_{\mu_c,\mu_s\rightarrow{0}} E[\mu n_c] = \bar{\mu}E[n_c],
 \end{equation}
 or equivalently,
 \begin{equation}\label{limiting_equation}
  \lim\limits_{\mu_c,\mu_s\rightarrow{0}} E[n_s n_c] = \rho_sE[n_c].
 \end{equation}
 \end{subequations}
 Combining equation (\ref{result3}) and (\ref{limiting_equation}), the lower bound of queue length is given by
 \begin{eqnarray}\label{lower_bound}
  L_1=\lim\limits_{\mu_c,\mu_s\rightarrow{0}} E[n_c] &=& \frac{\lambda_c}{\bar{\mu}-\lambda_c} \nonumber \\
             										 &=& \frac{\rho_c}{\rho_c-\rho_s}.
 \end{eqnarray}

 Intuitively, as $\mu_c \rightarrow 0$, the service rate becomes more static. Therefore, the P2P system can be regarded as a single server queue with a constant service rate in this limiting scenario. In fact, the formula $L_1$ is exactly the mean queue length of $M/M/1$ queue with arrival rate $\lambda_c$ and service rate $\bar{\mu}$. Since the mean queue length increases with the fluctuation of the service rate, $L_1$ becomes a lower bound of the mean queue length $L$ that corresponds to the limiting case $\sigma[\mu] \rightarrow 0$.\\

2. Upper bound $L_2 = \lim\limits_{\mu_c,\mu_s\rightarrow{\infty}}L = (1+\frac{\mu_c}{\mu_s})\frac{\lambda_c}{\bar{\mu}-\lambda_c}$.\\

For a fixed mean service rate $\bar{\mu} = \rho_s\mu_c = \frac{\lambda_s\mu_c}{\mu_s}$, the limiting case $\mu_c \rightarrow \infty$, while keeping both parameters $\frac{\mu_c}{\mu_s}$ and $\lambda_s$ constant, implies $\mu_s \rightarrow \infty$ and $\rho_s \rightarrow 0$. The physical meaning of the P2P system operated under this scenario can be interpreted as follows:

\begin{itemize}
 \item $\rho_s \rightarrow 0$ implies that the average number of servers is very small,
 \item $\mu_s \rightarrow \infty$ implies that the average lifetime of a server $\frac{1}{\mu_s}$ is very short,
 \item $\mu_c \rightarrow \infty$ implies that the capacity of each server, in terms of the number of jobs served per unit of time, is very large.
\end{itemize}

\begin{figure}[h]
\centering
\includegraphics[height=35mm]{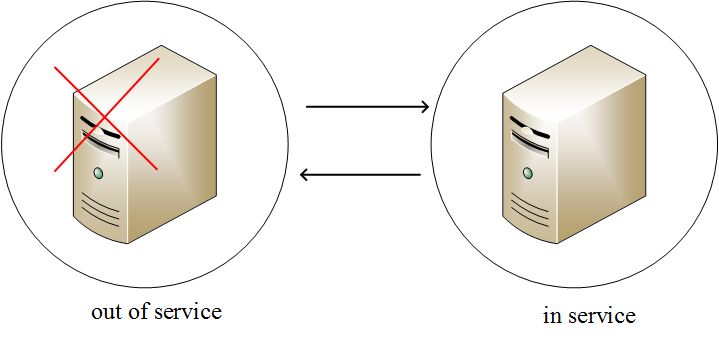}
\caption{The two-state extreme scenario of P2P systems.}\label{figure5}
\end{figure}

Fig. \ref{figure5} illustrates this extreme scenario. The probability that there are two or more servers co-existing in the system is negligible because the lifetime of a server is so short. The only server in the system has a very large service rate (capacity) but a very short lifetime; thus, the system sometimes could be out of servers. Therefore, the P2P system can be regarded as a queuing system with a two-state server, which is available in one state with a very large service rate $\mu_c$ but absent in the other state.

This extreme case of the P2P system can also be described by the distribution of the number of servers in a straightforward manner.  Since the number of servers $n_s$ follows Poisson distribution with parameter $\rho_s$, we have 
\begin{subequations}
 \begin{equation}
  Pr\{n_s=0\} = 1 - \rho_s + o(\rho_s^2),
 \end{equation}
 \begin{equation}
  Pr\{n_s=1\} = \rho_s + o(\rho_s^2),
 \end{equation}
 and
 \begin{equation}
  Pr\{n_s \geq 2\} = o(\rho_s^2).
 \end{equation}
\end{subequations}

 Since the probability that the number of servers is larger than 1 is negligible as $\rho_s$ approaches 0, which implies $p_{i,j} = 0$ for $j \geq 2$ . Therefore, the original P2P system will degenerate to a queueing model with a two-state server. The continuous time Markov chain depicted in Fig. \ref{figure6} illustrates the transition diagram of the two service states.

\begin{figure}[H]
\centering
\includegraphics[height=30mm]{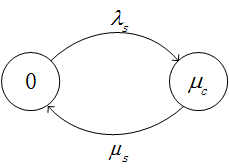}
\caption{The transition diagram of the two service states.}\label{figure6}
\end{figure}

A complete analysis of the queueing system with a two-state server is studied in \cite{HuangLiangGPK}\cite{Twostate2010}. The generating functions of the number of jobs at server state 0 and state 1 are respectively given by equation (8) in p.3533 of \cite{HuangLiangGPK} as follows:
\begin{subequations}
 \begin{eqnarray}\label{G0}
  G_0(z) &=& \sum_{i=0}^\infty p_{i,0} z^i \nonumber \\
  &=& \frac{(\frac{\lambda_s\mu_c}{\lambda_s+\mu_s}-\lambda_c)\mu_s}{z^2\lambda_c^2-z\lambda_c(\lambda_s + \mu_s +\lambda_c + \mu_c) + \mu_c(\lambda_s + \lambda_c)}, \nonumber \\
 \end{eqnarray}
 \begin{eqnarray}\label{G1}
  G_1(z) &=& \sum_{i=0}^\infty p_{i,1} z^i  \nonumber \\
  &=& - \frac{(\frac{\lambda_s\mu_c}{\lambda_s+\mu_s}-\lambda_c)(\lambda_s+\lambda_c-\lambda_c z)}{z^2\lambda_c^2-z\lambda_c(\lambda_s + \mu_s +\lambda_c + \mu_c) + \mu_c(\lambda_s + \lambda_c)}. \nonumber \\
 \end{eqnarray}
\end{subequations}
The upper bound of mean queue length $L_2$ can be readily obtained from the above generating functions and given as follows:
\begin{equation}
 L_2 = \lim\limits_{\mu_c,\mu_s\rightarrow{\infty}}G'_0(1) + G'_1(1). \nonumber \\
\end{equation}
First, taking the derivative of (\ref{G0}) and (\ref{G1}), we obtain
\begin{eqnarray}\label{G0+G1}
 &&G'_0(z) + G'_1(z) = \nonumber \\
&&\frac{(\frac{\lambda_s\mu_c}{\lambda_s+\mu_s}-\lambda_c)}{[z^2\lambda_c^2-z\lambda_c(\lambda_s + \mu_s +\lambda_c + \mu_c) + \mu_c(\lambda_s + \lambda_c)]^2} \times \nonumber \\
 &&\{ \lambda_c[z^2\lambda_c^2-z\lambda_c(\lambda_s + \mu_s +\lambda_c + \mu_c) + \mu_c(\lambda_s + \lambda_c)] + \nonumber \\
 &&(\lambda_s + \lambda_c - \lambda_c z - \mu_s)[2z\lambda_c^2-\lambda_c(\lambda_s + \mu_s +\lambda_c + \mu_c)]\}. \nonumber \\
\end{eqnarray}
Next, inserting $z=1$ into the above equation, we obtain
\begin{eqnarray}
 &&G'_0(1) + G'_1(1) = \nonumber \\
 &&\qquad -\frac{(\frac{\lambda_s\mu_c}{\lambda_s+\mu_s}-\lambda_c)[2\lambda_c^2\mu_s-\lambda_c\mu_s(\mu_s + \mu_c) + \lambda_s^2\lambda_c]}{[\mu_c\lambda_s-\lambda_c(\lambda_s+\mu_s)]^2}.\nonumber \\
\end{eqnarray}
Finally, taking the limit $\mu_c,\mu_s \rightarrow \infty$, the upper bound of the mean queue length is given by:
\begin{eqnarray}\label{upper_bound}
 L_2 &=& \lim\limits_{\mu_c,\mu_s\rightarrow{\infty}}G'_0(1) + G'_1(1) \nonumber \\
 &=& \frac{\lambda_c(\mu_s+\mu_c)}{\mu_c\lambda_s-\lambda_c\mu_s} \nonumber \\
 &=& (1+\frac{\mu_c}{\mu_s})\frac{\lambda_c}{\bar{\mu}-\lambda_c}.
\end{eqnarray}

Thus, the assertion of the theorem $L_1 \leq L \leq L_2$ is established.

\end{IEEEproof}

\begin{figure}[H]
\centering
\includegraphics[height=65mm]{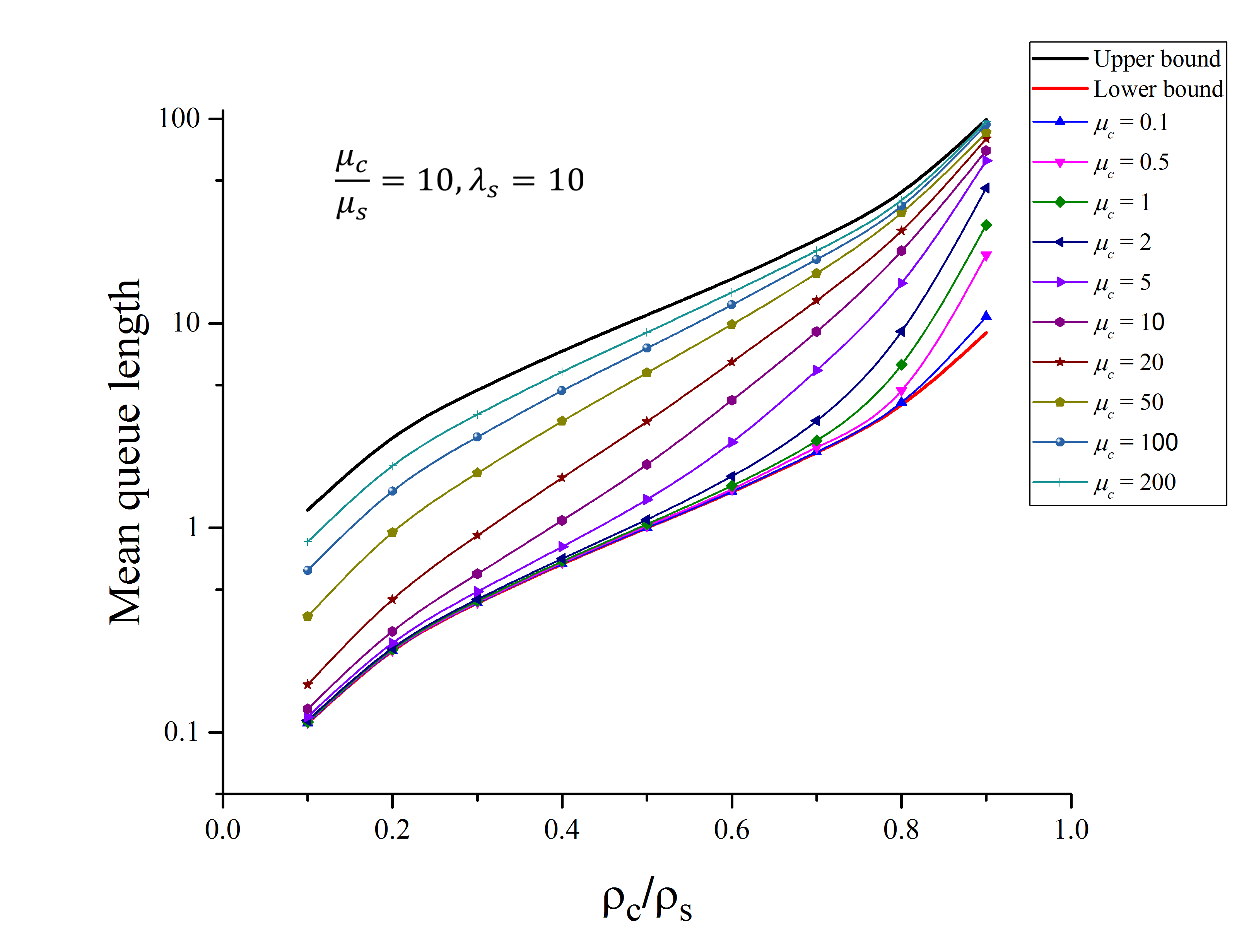}
\caption{Mean queue length $L$ is bounded by $L_1$ and $L_2$.}\label{figure7}
\end{figure}

Theorem 1 has been verified by simulation results, with fixed parameter $\frac{\mu_c}{\mu_s} = 10$ and $\lambda_s=10$, while $\lambda_c$ changes from 10 to 90. Fig. \ref{figure7} shows that the mean queue length $L$ approaches to the upper bound $L_2$ when $\mu_c \rightarrow \infty$, and the lower bound $L_1$ when $\mu_c \rightarrow 0$. 

Theorem \ref{theorem1} also offers the following stable condition of the P2P system:
\begin{subequations}\label{stable_condition}
 \begin{equation}
  \rho_c<\rho_s,
 \end{equation}
 or equivalently,
 \begin{equation}
  \lambda_c<\bar{\mu}.
 \end{equation}
\end{subequations}
This condition ensures that both the lower bound $L_1$ and the upper bound $L_2$ are positive and finite, which guarantee that the mean queue length $L$ is also positive and finite.

The lower bound conjecture on the mean queue length was based on an unproved postulation claimed in \cite{Taoyujournal} that the number of servers and the number of jobs are negatively correlated. The covariance of the number of servers and the number of jobs is given by
\begin{equation}\label{cov}
  Cov(n_s, n_c) = E[n_sn_c]-E[n_s]E[n_c].
\end{equation}
Substitute (\ref{result3}) into (\ref{cov}), we have
\begin{eqnarray}\label{25}
  Cov(n_s, n_c) &=&  \rho_cE[n_c]+\rho_c-\rho_sE[n_c] \nonumber \\
                &=& \rho_c - (\rho_s - \rho_c) L.
\end{eqnarray}
From (\ref{lower_bound}) and (\ref{25}) , we have
\begin{eqnarray}
  Cov(n_s, n_c) &=& \rho_c-(\rho_s-\rho_c)L \nonumber \\
                &\leq& \rho_c-(\rho_s-\rho_c)L_1 = 0,
\end{eqnarray}
which confirms the postulation and the equality holds when $\mu_c \rightarrow 0$.

\section{An Approximate P-K Formula of Mean Queue Length}\label{approximation}
This section derives a simple formula of the mean queue length for P2P systems. The derivation of the two bounds in the previous section indicates that the mean queue length of a P2P system is governed by the fluctuation of service rate. In traditional queueing analysis with constant service rate, we know that if the service rate is greater than the job arrival rate, then the queue length is always finite and the system is stable. On the other hand, if the service rate is smaller than the job arrival rate, then the queue length rapidly grows to infinity and the system becomes unstable.

In the P2P network, however, the total service rate varies with time. That means sometimes the service rate is greater than the job arrival rate, and other times it is smaller than the job arrival rate. According to Theorem \ref{theorem1}, the mean queue length achieves the lower bound when the service rate converges to a constant. In this case, this constant service rate must be larger than the job arrival rate to make the system stable.

As the service rate fluctuates, if the probability that the service rate is smaller than the job arrival rate becomes larger, then the system would perform worse than the lower bound. Thus, the fraction of time that the service rate is smaller than the job arrival rate determines how much the system would perform worse than the lower bound. In this section, from the two bounds $L_1$ and $L_2$ given in Theorem \ref{theorem1}, we derive the following approximate estimation of the mean queue length of P2P system:
\begin{equation}\label{first_L}
   L = (1+\frac{\mu_c}{\mu_s}\alpha)\frac{\lambda_c}{\bar{\mu}-\lambda_c},
\end{equation}
for some parameter $0 \leq \alpha \leq 1$.

\subsection{Mean queue length formula}\label{formula}

\begin{figure}[H]
\centering
\includegraphics[height=50mm]{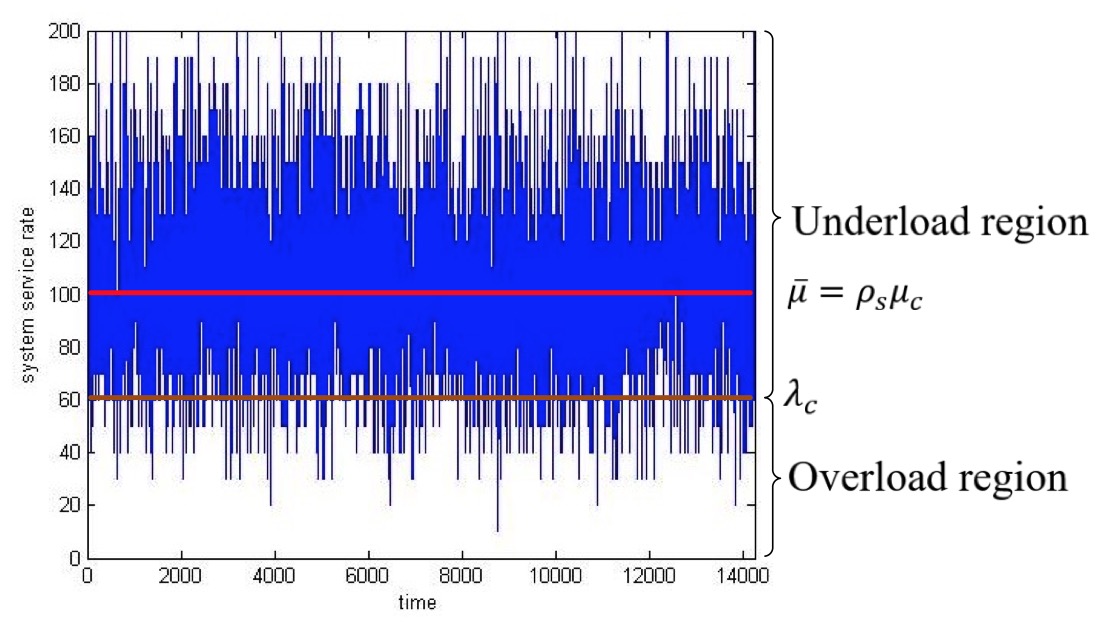}
\caption{Overload and underload regions.}\label{figure8}
\end{figure}

We first investigate the service rate fluctuation in time interval $[0,T]$. Since the number of servers $n_s(t)$ at time $t \in [0,T]$ may fluctuate around the mean $\rho_s$ as Fig. \ref{figure8} shows. Thus the time interval can be divided into two regions:
\begin{subequations}
 \begin{equation}
  T_{underload} = \{t|\mu(t)>\lambda_c, t \in[0,T]\},
 \end{equation}
 and
 \begin{equation}
  T_{overload} = \{t|\mu(t) \leq \lambda_c, t \in[0,T]\}.
 \end{equation}
\end{subequations}

The queue length is always finite in the underload region $T_{underload}$, because the service rate $\mu(t)$ exceeds the job arrival rate $\lambda_c$. In the overload region $T_{overload}$, however, the queue length may grow rapidly when the service rate $\mu(t)$ is lower than the job arrival rate $\lambda_c$. In steady state, we have 
\begin{subequations}
 \begin{equation}
  Pr\{\mu>\lambda_c\} = \lim\limits_{T\rightarrow{\infty}} \frac{T_{underload}}{T},
 \end{equation}
 and
 \begin{equation}\label{def_olPr}
  Pr\{\mu \leq \lambda_c\} = \lim\limits_{T\rightarrow{\infty}} \frac{T_{overload}}{T}.
 \end{equation}
\end{subequations}

We define the following notations used in the derivation of the mean queue length $L$:

\begin{itemize}
\item $L_{underload} = E[n_c|\mu > \lambda_c]$: conditional mean queue length of underload region,
\item $L_{overload} = E[n_c|\mu \leq \lambda_c]$: conditional mean queue length of overload region,
\item $a = Pr\{\mu \leq \lambda_c\}$: overload probability, and thus the underload probability equals $1-a$,
\item $b = Pr\{\mu \leq \bar{\mu}\}$: probability that service rate $\mu$ is smaller than mean service rate $\bar{\mu}$.
\end{itemize}

The mean queue length can be expressed as the combination of the conditional mean queue lengths as follows:
\begin{equation}\label{L_a}
  L = E[n_c] =  (1-a)L_{underload} + aL_{overload}.
\end{equation}

The above expression cannot help us evaluate the mean queue length because the two conditional mean values $L_{underload}$ and $L_{overload}$ are unknown. Since we know from Theorem \ref{theorem1} that the mean queue length $L$ is bounded by $L_1$ and $L_2$, therefore $L$ can be expressed as follows:
\begin{eqnarray}\label{L_b}
  L &=& (1-\alpha)L_1 + \alpha L_2 \nonumber \\
    &=& (1+\frac{\mu_c}{\mu_s}\alpha)\frac{\lambda_c}{\bar{\mu}-\lambda_c},
\end{eqnarray}
for some parameter $0 \leq \alpha \leq 1$. For a proper chosen parameter $\alpha$, the linear combination (\ref{L_b}) of $L_1$ and $L_2$, which is similar to expression (\ref{L_a}), can serve as a good approximation of the mean queue length $L$. 

Intuitively, the overload probability $a$ is a measurement that indicates how much the system performance is worse than the lower bound. When the overload probability $a$ is small, then the mean queue length $L$ should be close to the lower bound $L_1$. In particular, when $a \rightarrow 0$, the parameter $\alpha$ should also approaches 0. On the other hand, for larger overload probability $a$, the parameter $\alpha$ should also be larger, indicating that the mean queue length $L$ is closer to the upper bound $L_2$. Therefore, a proper choice of the parameter $\alpha$ is linearly proportional to the overload probability $a$:
\begin{equation}\label{assumption}
  \hat{\alpha} = ka.
\end{equation}

When the system becomes saturated as $\lambda_c$ approaches $\bar{\mu}$, or equivalently $\rho_c$ approaches $\rho_s$, we expect that the expression (\ref{L_b}) of mean queue length $L$ will reach the upper bound $L_2$. That is, the proportional constant $k$ can be determined by the following limiting condition:
\begin{equation}
  \lim\limits_{\lambda_c\rightarrow{\bar{\mu}}} \hat{\alpha} = k\lim\limits_{\lambda_c\rightarrow{\bar{\mu}}} Pr\{\mu \leq \lambda_c\} = kPr\{\mu \leq \bar{\mu}\} = 1,
\end{equation}
thus, we have
\begin{equation}\label{k}
  k = \frac{1}{Pr\{\mu \leq \bar{\mu}\}} >1.
\end{equation}
It follows from (\ref{assumption}) and (\ref{k}) that 
\begin{eqnarray}\label{alpha}
  \hat{\alpha} = \frac{a}{b} &=& \frac{Pr\{\mu \leq \lambda_c\}}{Pr\{\mu \leq \bar{\mu}\}}  \nonumber\\
                             &=& \frac{Pr\{ n_s \leq \rho_c\}}{Pr\{ n_s \leq \rho_s\}}.
\end{eqnarray}
 The above choice of the parameter $\alpha$ satisfies the required condition (\ref{first_L}), and the following mean queue length formula is readily obtained from (\ref{L_b}):

\textbf{Approximate P-K Formula for P2P Networks}
\begin{equation}\label{APK}
   L \cong (1+\frac{\mu_c}{\mu_s}\frac{a}{b} )\frac{\lambda_c}{\bar{\mu}-\lambda_c},
\end{equation}
where 
\begin{subequations}
 \begin{eqnarray}
  a = Pr\{\mu \leq \lambda_c\} &=& \sum_{i=0}^{\rho_c} \frac{\rho_s^i}{i!}e^{-\rho_s}  \nonumber\\
  &\triangleq& \sum_{i=0}^{\lfloor \rho_c \rfloor} \frac{\rho_s^i}{i!}e^{-\rho_s} + (\rho_c-\lfloor \rho_c \rfloor) \frac{\rho_s^{\rho_s}}{\Gamma(\rho_s)}e^{-\rho_s},\nonumber
 \end{eqnarray}
 and
 \begin{eqnarray}
 b = Pr\{\mu \leq \bar{\mu}\} &=& \sum_{i=0}^{\rho_s} \frac{\rho_s^i}{i!}e^{-\rho_s}  \nonumber \\
 &\triangleq& \sum_{i=0}^{\lfloor \rho_s \rfloor} \frac{\rho_s^i}{i!}e^{-\rho_s} + (\rho_s-\lfloor \rho_s \rfloor) \frac{\rho_s^{\rho_s}}{\Gamma(\rho_s)}e^{-\rho_s}.\nonumber
 \end{eqnarray}
\end{subequations}
This formula exhibits the strong correlation among the service times of different jobs. Instead of the first two moments of the service time in the P-K formula of M/G/1 queue, the determination of the mean queue length of P2P networks needs the entire distribution of the service rate.

\begin{figure}[H]
	\center
    \subfigure[$\mu_c=10$]{
    \includegraphics[height=60mm]{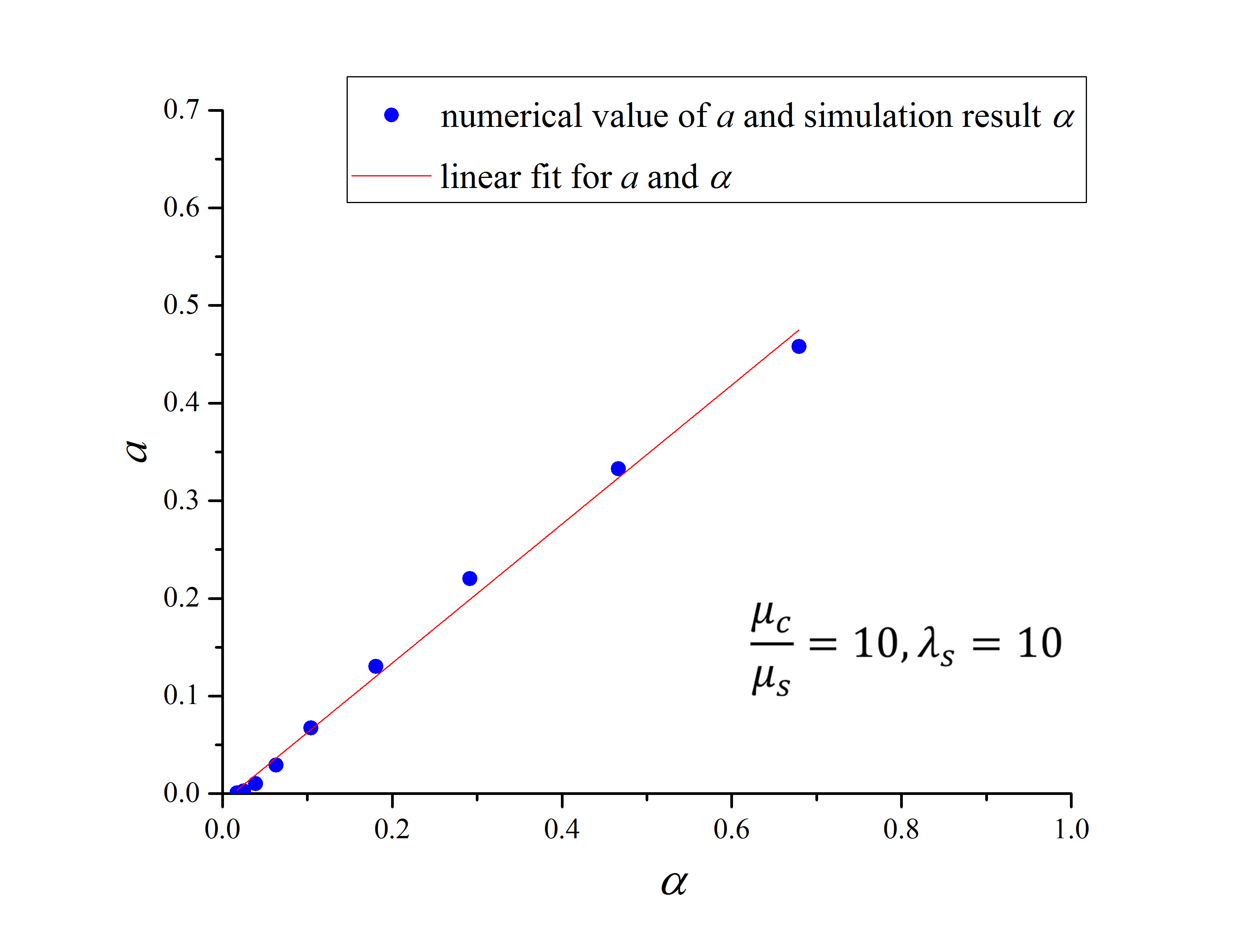}}
    \subfigure[$\mu_c=50$]{
    \includegraphics[height=60mm]{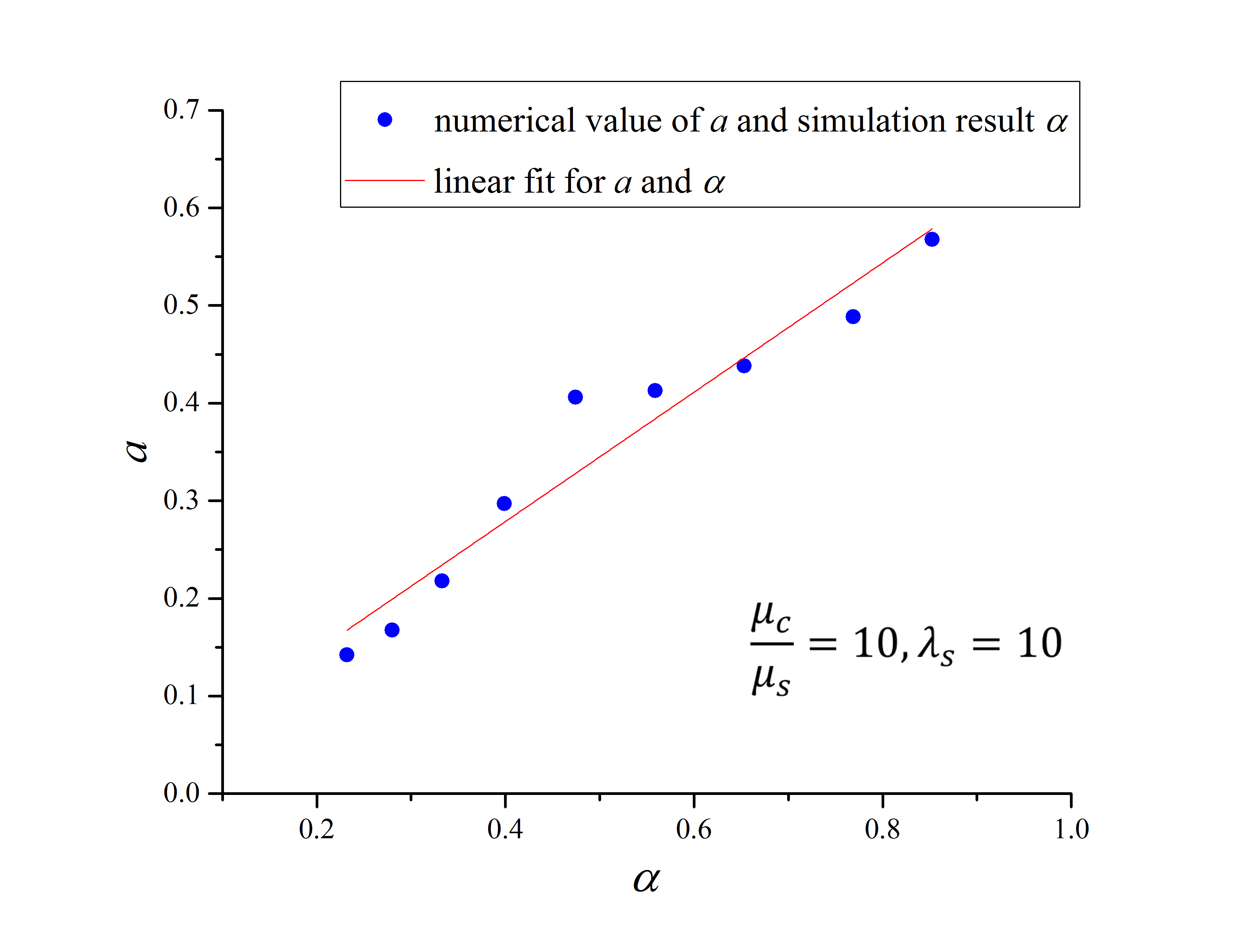}}
    \caption{Overload probability $a$ versus parameter $\alpha$.}\label{figure9}
\end{figure}

The choice of the parameter $\alpha$ is based on the assumption that it is linearly proportional to $a$. This assumption is confirmed by simulations. As Fig. \ref{figure9} shows, the simulation results demonstrate that the parameter $\alpha$ in (\ref{L_b}) given by $\alpha = \frac{L-L_1}{L_2-L_1}$ is indeed linearly proportional to the overload probability $a$.

Our simulation results also show that the following relation
 \begin{equation}\label{overload}
  L_{overload} = L_2
 \end{equation}
 holds in most cases. Assuming that the relation $L_{overload} = L_2$  holds, substitute (\ref{L_b}) into (\ref{L_a}), we have
\begin{equation}\label{L_under}
  L_{underload} = (1-\alpha')L_1 + \alpha'L_2,
\end{equation}
where $\alpha' = \frac{\alpha-a}{1-a}$. Since we know $\hat{\alpha}=ka>a$, which ensures that the parameter $\alpha'$ also follows $0 \leq \alpha'<1$, thus the conditional mean queue length $L_{underload}$ can be expressed by:
\begin{eqnarray}\label{underload}
 L_{underload} &\cong& (1+\frac{\mu_c}{\mu_s}\alpha')\frac{\lambda_c}{\bar{\mu}-\lambda_c} \nonumber \\
                &=& (1+\frac{\mu_c}{\mu_s}\frac{a(1-b)}{b(1-a)})\frac{\lambda_c}{\bar{\mu}-\lambda_c}.
\end{eqnarray}

\begin{figure}[H]
	\center
    \subfigure[$L_{overload}$]{
    \includegraphics[height=60mm]{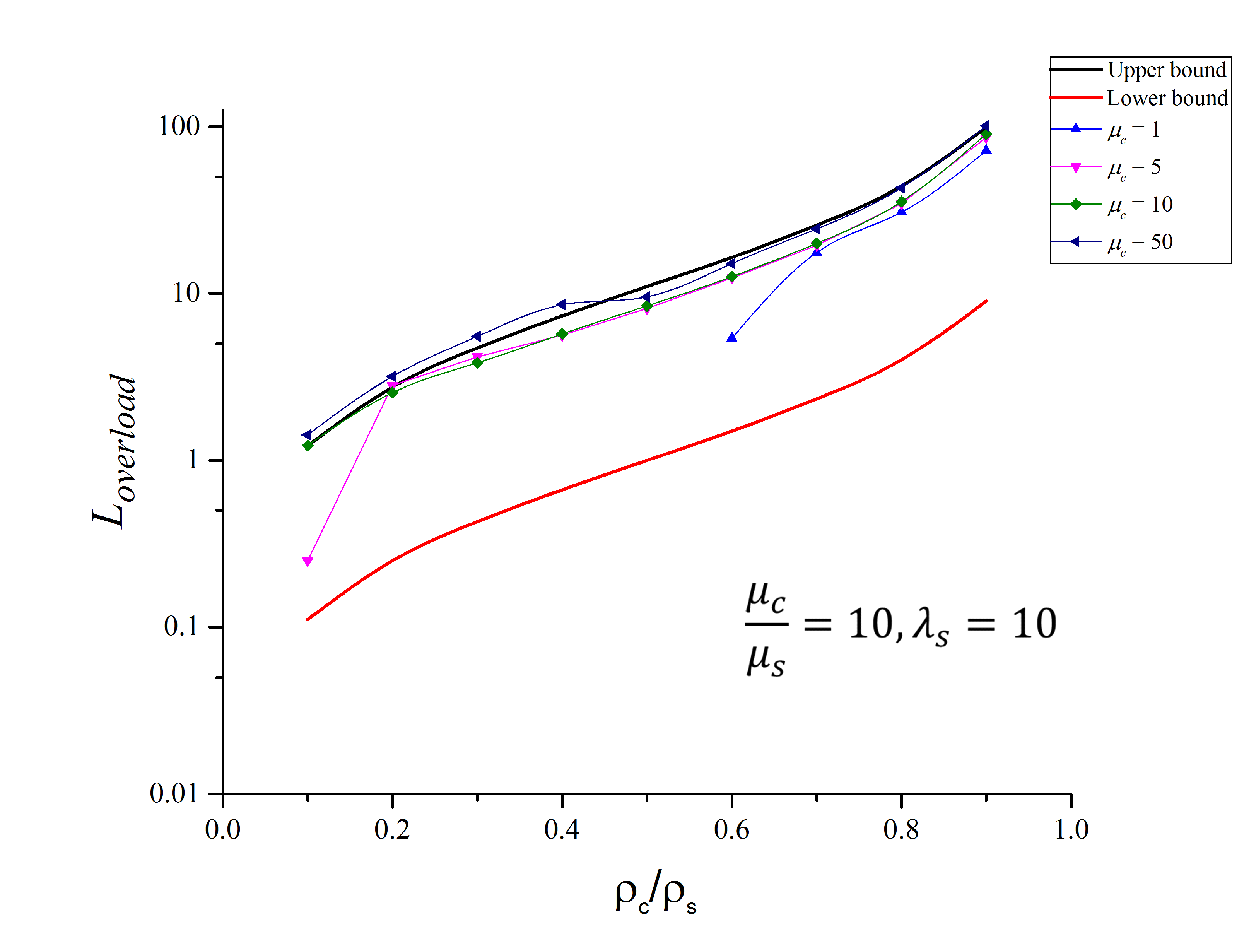}}
    \subfigure[Overload probability $a$]{
    \includegraphics[height=60mm]{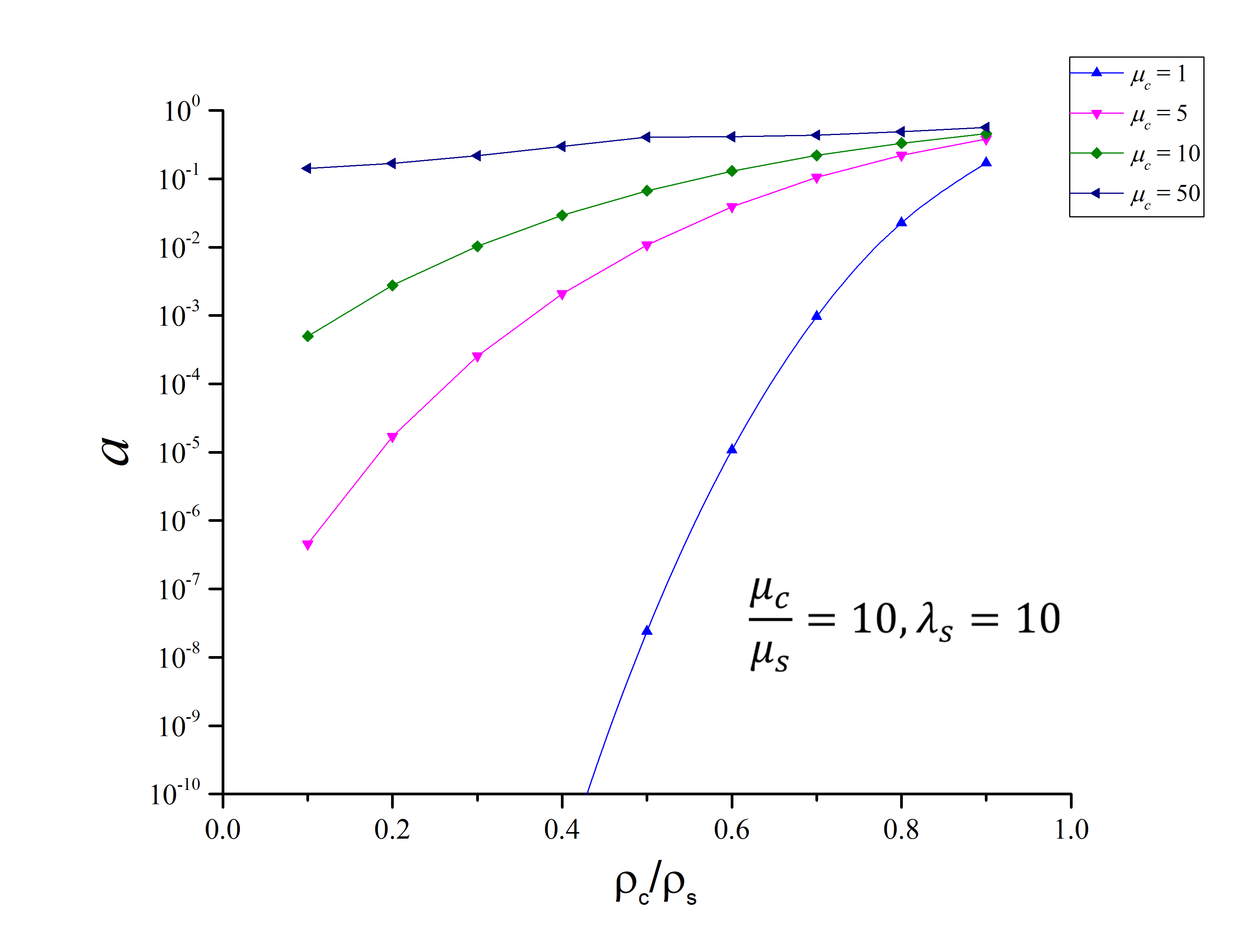}}
    \caption{Mean queue length in overload region $L_{overload}$ and overload probability $a$.}\label{figure10}
\end{figure}

 It should be noted that if the assumption that the relation $L_{overload} = L_2$ holds, then the expression (\ref{L_a}) of the mean queue length $L$ is exactly the combination of (\ref{overload}) and (\ref{underload}). However, this assumption sometimes may fail to hold when unit service rate $\mu_c$ and job arrival rate $\lambda_c$ are both small. As the simulation results displayed in Fig. \ref{figure10} demonstrate, the mean queue length in the overload region $L_{overload}$ is far below the upper bound $L_2$ when $\mu_c=1$ and $\frac{\rho_c}{\rho_s}$ is less than 0.6 or $\mu_c=5$ and $\frac{\rho_c}{\rho_s}$ is less than 0.2. However, we also notice that the system rarely comes across to the overload region under this condition. As a matter of fact, our simulation could not collect any data in the overload region if $L_{overload}$ is much less than $L_2$. The reason is that the system does not stay in the overload region long enough to reach the upper bound $L_2$ if the fraction of overload time $T_{overload}$ is too small compared to that of $T_{underload}$. As a result, the overload probability $a = Pr\{\mu \leq \lambda_c\}$, defined by (\ref{def_olPr}), would be very small. Hence, the error introduced by the discrepancy between $L_{overload}$ and $L_2$ is negligible. That is, the formula (\ref{APK}) remains a sound estimation of mean queue length $L$ even if the relation $L_{overload} = L_2$ fails to hold.

\begin{figure*}[h]
	\center
    \subfigure[$\mu_c=1$]{
    \includegraphics[width=80mm]{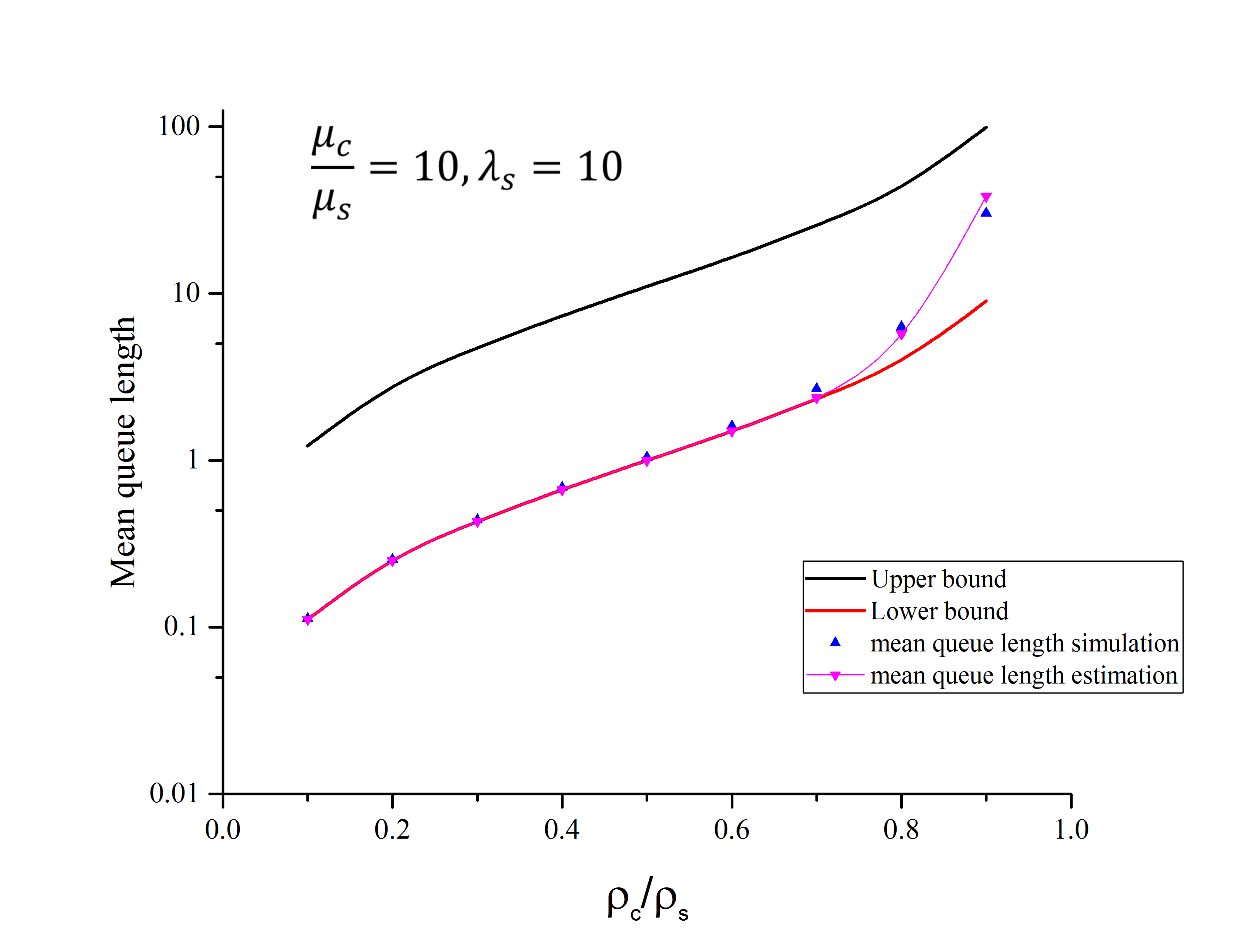}}
    \hspace{5mm}
    \subfigure[$\mu_c=5$]{
    \includegraphics[width=80mm]{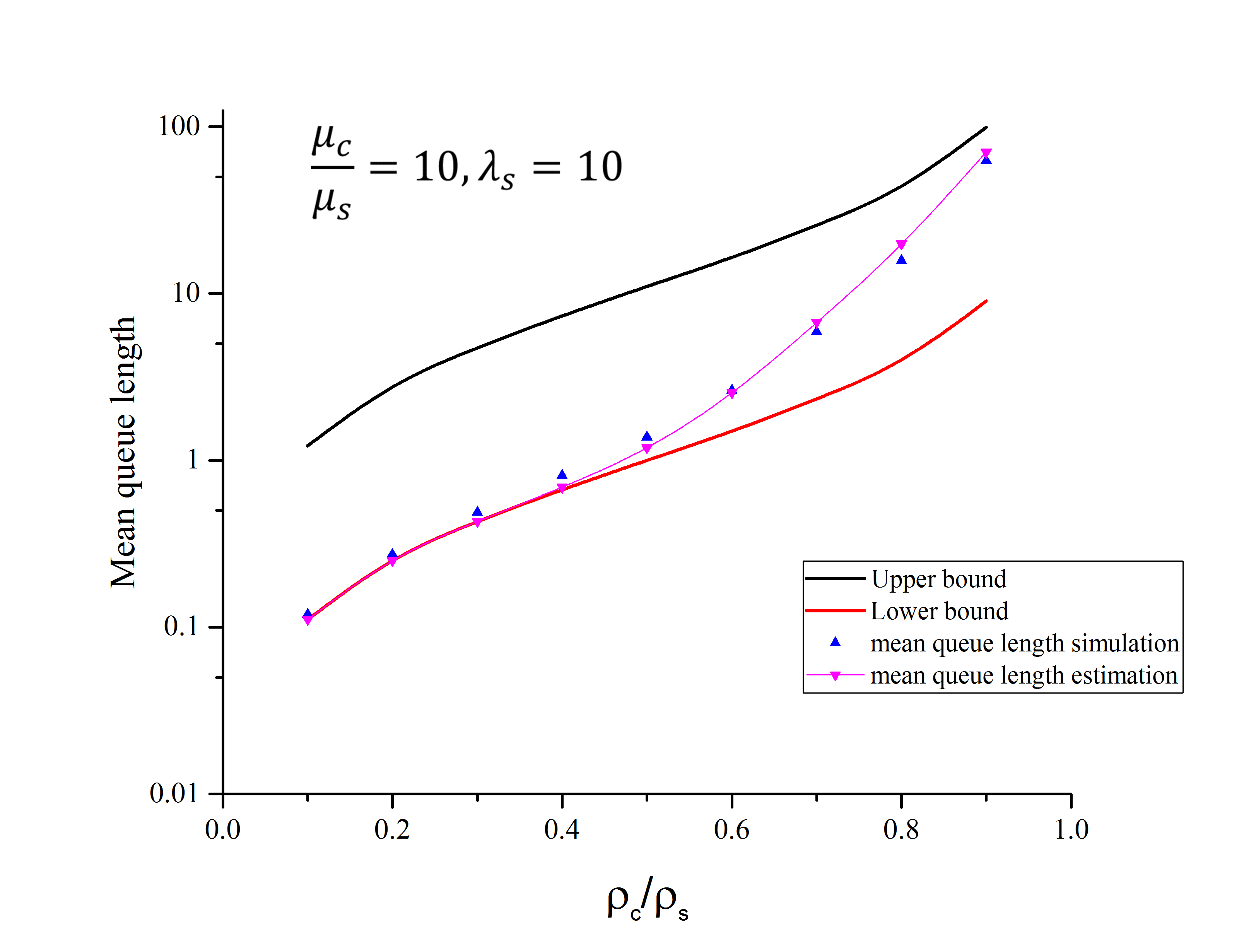}}
    \subfigure[$\mu_c=10$]{
    \includegraphics[width=80mm]{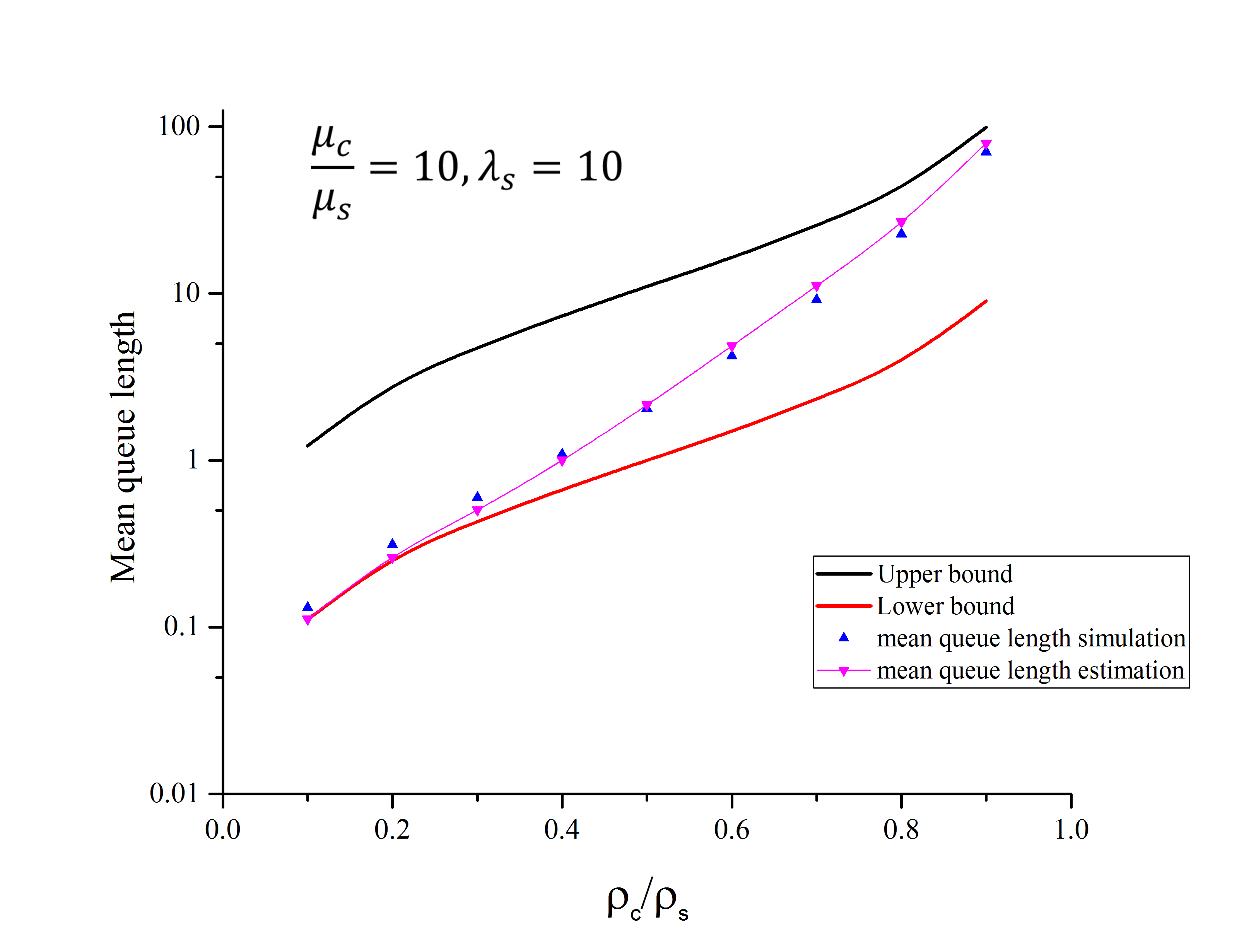}}
    \hspace{5mm}
    \subfigure[$\mu_c=50$]{
    \includegraphics[width=80mm]{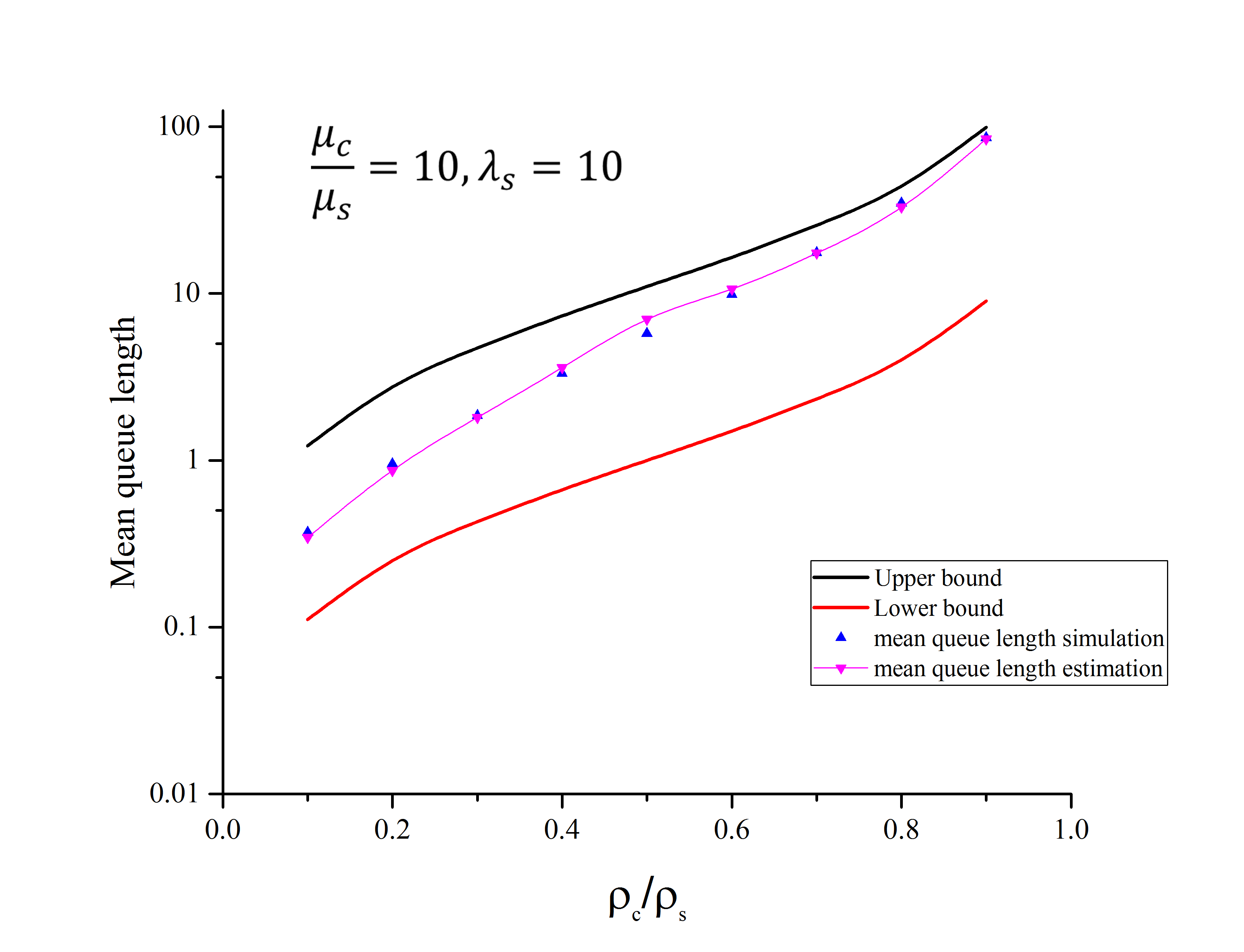}}
    \caption{Simulation and approximation results of mean queue lengths.}\label{figure11}
\end{figure*}

The robustness of the mean queue length formula (\ref{APK}) is demonstrated by simulation results displayed in Fig. \ref{figure11}, in which we use parameter $\frac{\mu_c}{\mu_s} = 10$, $\lambda_s=10$ and $\lambda_c$ changes from 10 to 90, therefore $\frac{\rho_c}{\rho_s}=\frac{\lambda_c\mu_s}{\lambda_s\mu_c}$ changes from 0.1 to 0.9. The mean queue lengths estimated by formula (\ref{APK}) agree with the simulation results in all cases we considered, even when $L_{overload}$ is far below the upper bound, since in this case, the probability of overload is below $10^{-4}$. For example, Fig. \ref{figure10} shows that the $L_{overload}$ is much smaller than the upper bound $L_2$ when $\frac{\rho_c}{\rho_s}$ is below 0.6, and $\mu_c=1$. However, as Fig. \ref{figure11}(a) shows, with the same set of parameters, the overall mean queue length $L$ estimated by formula (\ref{APK}) still fits very well with the simulation result.

\subsection{Verifications of limiting cases}\label{limiting_cases}

In this subsection, we verify all limiting cases of the system behavior. The results confirm that all limiting cases are consistent with the mean queue length formula (\ref{APK}).

1. Lower bound

In the limiting case $\mu_c \rightarrow 0$ for a fixed mean service rate $\bar{\mu}$, the service rate becomes a constant, that is $\mu(t) = \bar{\mu}$ for all $t$. Since the stable condition $\rho_c<\rho_s$ holds if and only if $\lambda_c \in [0,\bar{\mu})$, we therefore must have $Pr\{\mu \leq \lambda_c\}=0$. In this case $\alpha = \frac{Pr\{\mu \leq \lambda_c\}}{Pr\{\mu \leq \bar{\mu}\}} =0$, then the mean queue length reduces to $L_1 = \frac{\lambda_c}{\bar{\mu}-\lambda_c}$.

2. Upper bound

In the limiting case $\mu_c \rightarrow \infty$ for a fixed mean service rate $\bar{\mu}$, as Fig. \ref{figure6} shows, the server arrival process can be described as a two-state Markov chain. The number of servers in the system is either $n_s = 0$ or $n_s = 1$, and the corresponding service rate is either $\mu = 0$ or $\mu = \mu_c$, while the mean service rate $\bar{\mu} = \mu_c Pr\{n_s = 1\}<\mu_c$. Since the stable condition (\ref{stable_condition}) implies $\lambda_c \in [0,\bar{\mu})$, then we must have  
\begin{equation}
Pr\{\mu \leq \lambda_c\}= Pr\{\mu \leq \bar{\mu}\} = Pr\{n_s = 0\}.
\end{equation}
It follows that $\alpha = \frac{Pr\{\mu \leq \lambda_c\}}{Pr\{\mu \leq \bar{\mu}\}} =1$, in which case (\ref{APK}) reduces to the upper bound $L_2=(1+\frac{\mu_c}{\mu_s}\alpha)\frac{\lambda_c}{\bar{\mu}-\lambda_c}$.

3. $\lambda_c \rightarrow 0$

When the job arrival rate $\lambda_c$ is very small, the overload probability can be interpreted as 
\begin{equation}
\lim\limits_{\lambda_c\rightarrow{0}} a = Pr\{n_s = 0\} = e^{-\rho_s} \nonumber. 
\end{equation}
There are two subcases of interest:
\begin{itemize}
\item If $\mu_s$ is relatively small, then the average number of servers $\rho_s$ is very large, and we have
\begin{equation}
\alpha = \frac{a}{b} = \frac{e^{-\rho_s}}{Pr\{\mu \leq \bar{\mu}\}} \rightarrow \frac{0}{1} = 0,
\end{equation}
which implies that the mean queue length approaches the lower bound $L_1$.

\item When $\mu_s \rightarrow \infty$, then the average number of servers $\rho_s$ is very small, we have
\begin{equation}
\alpha =\frac{a}{b} =  \frac{e^{-\rho_s}}{Pr\{\mu \leq \bar{\mu}\}} \rightarrow \frac{e^{-\rho_s}}{e^{-\rho_s}} =1,
\end{equation}
which implies that the mean queue length approaches the upper bound $L_2$.
\end{itemize}

Thus, when the offer load is small, the system performance is mainly determined by the server process. This analysis generally agrees with the simulation result shown in Fig. \ref{figure7}.

4. $\lambda_c \rightarrow \bar{\mu}$

In this case, it is obvious that $\alpha = \frac{Pr\{\mu \leq \lambda_c\}}{Pr\{\mu \leq \bar{\mu}\}} \rightarrow 1$, which implies that the mean queue length of the system reaches the upper bound when the offer load is close to saturation. In Fig. \ref{figure7}, the simulation results also show that the mean queue length approaches the upper bound when $\frac{\rho_c}{\rho_s} \rightarrow 1$.

\section{Conclusion}\label{conclusion}
In this paper, the P2P networks are modeled by a two-dimensional Markov chain, in which jobs and servers arrive and depart randomly. However, it is mathematically intractable to obtain a closed-form solution of the balance equations of this Markov chain. Inspired by the P-K formula for M/G/1 queue, we establish a connection between the fluctuation of the service rate and the mean queue length. For two extreme cases of service rates, we prove the conjecture on the lower bound and upper bound of the mean queue length postulated by Taoyu Li et al. in \cite{Taoyujournal}.

Furthermore, we provide a simple formula to estimate the mean queue length. The accuracy of our approximation has been verified by extensive simulation studies with different parameters. Furthermore, all limiting cases of the system behavior we checked completely agree with the predictions made by our formula. Thus, this formula could serve as a useful tool in the study of the performance of P2P systems.

Using the queueing system with varying service rate to model complicated real network applications becomes more and more difficult. Except in some simple special cases, most of them are not solvable by using traditional queueing analysis. We expect that the approach developed in this paper will shed some light on the queueing model with variable service rate, and the proposed methodology can be extended and applied to many other research fields, such as mobile cloud computing or energy efficient Ethernet.

\bibliography{IEEEabrv,P2P}
\end{document}